\documentclass[11pt, a4paper, logo, onecolumn, copyright,numbers]{googledeepmind}
\pdftrailerid{redacted}

\usepackage{kantlipsum, lipsum}
\usepackage{dsfont}
\usepackage{dm-colors}

\usepackage{algorithm}
\usepackage{algpseudocode}
\usepackage{adjustbox}
\usepackage{multirow}
\usepackage{nicefrac}
\usepackage{cleveref}
\usepackage{soul}

\usepackage{titlesec}
\titleformat{\section}
  {\normalfont\bfseries\normalsize} 
  {\thesection.}{0.5em}{#1}

\crefformat{section}{\S#2#1#3} 
\crefformat{subsection}{\S#2#1#3}
\crefformat{subsubsection}{\S#2#1#3}

\newcolumntype{R}[2]{%
    >{\adjustbox{angle=#1,lap=\width-(#2)}\bgroup}%
    l%
    <{\egroup}%
}

\usepackage[sort&compress]{natbib} 
\usepackage{comment}

\usepackage{pifont}
\makeatletter
\renewcommand\@cite[2]{[#1]\if@tempswa , #2\fi}
\makeatother

\usepackage{subcaption}
\usepackage{url}

\graphicspath{{figures/}}

\title{\Large{Beyond Moore’s Law: Harnessing the Redshift of Generative AI with Effective Hardware-Software Co-Design}}

\correspondingauthor{ayazdan@google.com}

\reportnumber{} 
\renewcommand{\today}{2025-04-08} 

\author{Amir Yazdanbakhsh\\Google DeepMind}

\begin{abstract}
For decades, Moore’s Law has served as a steadfast pillar in computer architecture and system design, promoting a clear abstraction between hardware and software. 
This traditional Moore’s computing paradigm has deepened the rift between the two, enabling software developers to achieve near-exponential performance gains often without needing to delve deeply into hardware-specific optimizations.
Yet today, Moore’s Law---with its once relentless performance gains now diminished to incremental improvements---faces inevitable physical barriers. 
This stagnation necessitates a reevaluation of the conventional system design philosophy.
The traditional decoupled system design philosophy, which maintains strict abstractions between hardware and software, is increasingly obsolete. 
The once-clear boundary between software and hardware is rapidly dissolving, replaced by co-design.
It is imperative for the computing community to intensify its commitment to hardware-software co-design~\cite{hwco}, elevating system abstractions to first-class citizens~\cite{abst} and reimagining design principles to satisfy the insatiable appetite of modern computing.
Hardware-software co-design is not a recent innovation.
Although pinpointing the exact inception is challenging, co-design likely emerged as an intuitive response to system complexity and cost, aiming to balance trade-offs between silicon and code. 
Historical records indicate that the explicit term “hardware-software co-design” gained popularity in the 1990s, even though its underlying principles date back much further, tracing back to the design of the ARPANET’s Interface Processors (IMPs)~\cite{wikipedia:IMP} in 1969.  
Since then, co-design philosophy evolved alongside Moore's computing paradigm, albeit with less emphasis and perhaps less necessity in earlier times. 
To illustrate its historical evolution, I classify its development into five relatively distinct “epochs”. 
This post also highlights the growing influence of the architecture community in interdisciplinary teams—particularly alongside ML researchers—and explores why current co-design paradigms are struggling in today’s computing landscape.
Additionally, I will examine the concept of the “hardware lottery”~\cite{hwlottery} and explore directions to mitigate its constraining influence on the next era of computing innovation.
\end{abstract}

\begin{document}

\maketitle

\makeatletter
\fancyhead[C]{\color{black}\normalfont\fontsize{10}{12}\selectfont Beyond Moore's Law: Harnessing the Redshift of Generative AI with Effective Hardware-Software Co-Design}
\makeatother

\section{From Moore’s Computing to Co-design for Intelligence}

\subsection{Epoch 1 - The Primordial Genesis of Modern Computing}
Between 1960 and 1980, the foundations of modern computing were laid amid a whirlwind of innovation and experimentation, a veritable Big Bang of technology. 
Although chaotic, this fertile period saw engineers and researchers exploring both decoupled and co-design philosophies. 
Early computing saw the birth of fundamental concepts and architectures that would later coalesce into the organized systems we rely on today.
This epoch, marked by rapid technological breakthroughs and a relative lack of established norms, set the stage for subsequent advancements in computing.
While formal co-design processes had not yet been established, early research prototypes like the MIT Tagged Token Dataflow Architecture~\cite{ttd} and the Manchester Dataflow Machine~\cite{manchester} embodied its underlying principles. Their success hinged on aligning hardware capabilities with software techniques to exploit fine-grained parallelism, demonstrating that even in its infancy, integrated hardware-software strategies could yield performance benefits.

On the commercial side, IBM 7030 (aka. Stretch)~\cite{IBM7030}, the first transistorized supercomputer, is often cited as an early example of a commercial co-design approach. 
The Stretch project brought hardware engineers, software developers, and system architects together from the very beginning. 
They recognized that achieving ambitious performance targets beyond traditional limits required a tightly integrated design process in which the ISA, microarchitecture, compiler technology, and even programming models were developed in tandem. 
Despite its ambitious goals, the Stretch system---though commercially unsuccessful---introduced innovations such as instruction pipelining, memory interleaving, prefetch, and specialized instructions. 
These innovations were not just added on top of a conventional design but were co-developed with the software in mind. This co-design methodology set an important precedent, demonstrating that breakthroughs often demand a full-stack, collaborative mindset. 
Its failure can be attributed to multiple factors, including overambitious goals, high cost and production issues, and market readiness.
Notably, the integrated co-design approach, while innovative, also introduced substantial technical complexity.
Developing hardware, software, and compilers concurrently was perhaps the right design philosophy at the time, but technological non-readiness and unforeseen challenges led to delays and difficulties in achieving stable, reliable performance.

After the Stretch project, IBM introduced the System/360 mainframe, representing a radical departure from previous approaches. 
While it still required close collaboration between hardware and software teams, it pioneered the abstraction of an “architecture” layer by introducing a standardized ISA that decoupled the software interface from the underlying hardware implementation (microarchitecture). 
This departure from the tightly integrated co-design approach, favoring modularity over complete integration, enabled different S/360 models to deliver varying capabilities while preserving software compatibility, thereby laying the foundation for modern computer architecture principles~\cite{hennessy2017computer}.

Two other commercially unsuccessful supercomputers from this period were the CDC 8600 and STAR projects.
Though innovative, these machines struggled primarily due to scalability challenges and the sheer complexity of coordinating hardware and software innovations under extremely aggressive (and perhaps unrealistic!) performance targets. 
Critically, both machines overlooked a foundational tenet of co-design: the importance of aligning design decisions with the prevailing computational patterns of applications at the time. 
By relying excessively on specialized vector units at the expense of scalar performance, these systems inadvertently amplified the overhead associated with vector setup, causing this overhead to dominate the execution time and ultimately degrade overall performance.

Learning from these oversights, Seymour Cray revisited the core principles of co-design when developing Cray-1. He carefully balanced powerful vector processing with robust scalar performance and an efficient register-based architecture. By aligning hardware design closely with compiler and runtime systems, Cray created a machine tailored precisely to real-world workloads. This meticulous, balanced co-design was pivotal in the Cray-1 commercial success, establishing it a milestone in computing history.

The contrasting fates of these machines highlight a critical lesson: true co-design is not merely about pushing hardware specialization to extremes; rather, it involves a thoughtful, balanced integration across hardware, software, and real-world workloads. 
Even today, this principle remains foundational, showing that successful innovations depend not solely on raw performance but also on thoughtful, integrated, and application-driven co-design~\cite{gemmini2021,jouppi2017in,chen2016eyeriss,tandem}. 
\subsection{Epoch 2 –  Hardware/Software Co-design for General-Purpose Computing} 
Rapid technological advances in this epoch allowed many of the visions from the primordial era to materialize.  Research on dataflow architectures continued, characterized by principled and tightly integrated  hardware-software co-design. The Monsoon architecture~\cite{monsoon}, emerging from earlier dataflow concepts~\cite{ttd}, exemplified a disciplined yet pragmatic approach to hardware-software co-design. Rather than inheriting all the complexity of previous dataflow designs, Monsoon significantly simplified dynamic dataflow execution. Its explicit token-store model and compiler-managed resource allocation provided an early illustration of how balancing complexity between hardware and software could yield practical and efficient implementations. 

Despite these promising innovations, general-purpose dataflow architectures ultimately did not gain widespread commercial traction, perhaps due to technological complexity and the rapidly increasing transistor budgets promised by Moore’s Law. With more transistors readily available, simpler designs could achieve substantial performance gains without incurring the extensive software complexity and overhead associated with dataflow machines. 
Additionally, market inertia may have also played a critical role, posing an additional barrier due to the existing software ecosystem's strong preference for von Neumann architectures. 

On the commercial front, the concept of co-design for general-purpose computers gained significant momentum with the advent of the VLIW architecture by Josh Fisher~\cite{vliw}.
VLIW was arguably the first true general-purpose architecture explicitly leveraging co-design principles, albeit somewhat imbalanced. 
Its architecture relied heavily—some might even argue, this was its Achilles' heel—on compiler technology to exploit instruction-level parallelism (ILP). 
Encouraged by its technical potential, Fisher founded Multiflow, a company dedicated to commercializing VLIW architectures using trace scheduling. 
Although Multiflow's commercial success was short-lived, it convincingly demonstrated the practicality of the VLIW design philosophy, despite initial skepticism. Multiflow’s achievements, though brief commercially, had significant technical influence on the superscalar movement.  After this unfortunate and slow acceptance of VLIW design in general-purpose computing, its design style has become a force in high-performance embedded systems. 

Indeed, these early experiences underscore how co-design philosophy evolved during this epoch. Initially focused heavily on theoretical elegance, the community progressively recognized that effective co-design demands a careful balance—complex enough to harness real-world benefits yet simple enough to remain practical. This epoch revealed that successful co-design in general-purpose computing requires disciplined hardware-software collaboration, thoughtful simplification, and sensitivity to practical constraints.
\subsection{Epoch 3 – The Golden Age of Moore’s Law and Scaling}
The 1990s marked a period of technological breakthroughs and mature design methodologies across the computing stack, exemplifying the peak visibility and impact of Moore’s Law~\cite{moore1965cramming}. 
The rapid transistor scaling incentivized the continued reliance on clear ISA abstractions, placing conventional, highly integrated hardware-software co-design for general-purpose computing into temporary hibernation. 
The community’s focus shifted toward architectural innovations rather than radically new co-design paradigms. 
As a result, traditional co-design was pushed into specialized niches, such as low-power embedded systems and real-time computing—domains characterized by stringent constraints and more integrated ecosystems.
Meanwhile, the widespread adoption of personal computing and the explosive growth of the internet further reinforced the dominance of general-purpose computing during this period.

A confluence of architectural innovations, mature and well-thought ISA abstractions, and stringent demand for backward compatibility—perhaps largely a consequence of widespread personal computing—catalyzed the success of superscalar processors. Notably, the commercial success of superscalar processors owed much to prior architectural breakthroughs, particularly out-of-order execution enabled by the exquisitely engineered Tomasulo’s algorithm~\cite{tomasulo}, first demonstrated in IBM System/360 Model 91. Nevertheless, escalating complexity in dependency checking, the overhead of register renaming circuitry, and diminishing returns from ILP motivated the exploration of alternative co-design architectures, such as EPIC~\cite{epic}, SMT, and multi-core processors.

The EPIC architecture, arguably the cousin of VLIW, leaned heavily on compiler support to explicitly identify parallelism and generate optimal instruction schedules. However, the lack of sufficient maturity in compiler technology at the time made this heavy reliance burdensome, posing substantial challenges for compiler developers. Combined with the requirement of an entirely new ISA, which significantly disrupted the existing x86-based software ecosystem, EPIC faced considerable market resistance. These factors, along with overly optimistic expectations and unmet performance promises, ultimately led to EPIC's commercial downfall. The EPIC experience provides a fundamental co-design lesson: successful architecture innovations require a thorough understanding of the target audience and sensitivity to practical software stack constraints.

Meanwhile, SMT and multicore architectures rose steadily, dealing a decisive blow to EPIC’s prospects. SMT architectures leveraged transistor abundance to better utilize processor resources with minimal added hardware complexity, while multicore processors emerged in response to physical constraints that limited further improvements in single-core frequency scaling.

Additionally, projects such as the Stanford DASH Multiprocessors~\cite{dash} demonstrated that directory-based cache coherence was a practical and effective approach to scaling shared-memory multiprocessors. DASH’s pragmatic, balanced co-design significantly influenced subsequent parallel computing architectures. Another noteworthy co-design exploration during this epoch was the introduction of PRISC (Programmable Reduced Instruction Set Computers)~\cite{prisc}, which offered an alternative approach emphasizing hardware adaptability through compiler-managed configurable logic. 

More broadly, this epoch explored diverse co-design strategies~\cite{registerwindow}, seeking a balanced middle ground rather than radically favoring either hardware- or software-centric solutions. For example, various compiler and architectural innovations were introduced to address branch-related performance bottlenecks in non-parallel code~\cite{branch1,branch2}.  Ultimately, this epoch offers valuable co-design lessons: successful architectures must strike a careful balance between hardware and software complexity, deliver robust performance improvements while preserving essential compatibility with legacy systems, and deeply understand the target ecosystems and workloads. These principles continue to shape the trajectory of modern computing design.
\subsection{Epoch 4 – Co-design Strikes Back: Domain-Specific Accelerators–The Phoenix of Modern Computing} 
By the late 2000s, Moore’s Law enabled chips with billions of transistors, fostering heterogeneous computing on a single die.
However, co-design received less emphasis for general-purpose computing.
Early signals of Moore’s Law saturation began emerging in the mid-2000s.
Around the same time, the concept of dark silicon challenged conventional hardware-software design philosophies. 
These shifts led to a noticeable trend toward domain-specific architectures. 
As a result, system development became increasingly specialized, with designs optimized explicitly for particular applications.
Faced with the slowdown of Moore’s Law and a widening gap between general-purpose performance and the orders-of-magnitude speed-ups needed for scientific simulations, the Anton machine~\cite{anton} was born in 2008. 
By achieving an unprecedented 500× performance improvement compared to commodity hardware available at the time, Anton vividly demonstrated the power of a reimagined co-design mindset in addressing specialized, computationally demanding tasks in scientific research, setting a precedent for future DSAs.

Google’s Tensor Processing Unit (TPU)~\cite{jouppi2017in}, another prime example of co-design, achieved 30×–80× higher TOPS/Watt on deep neural network workloads by co-designing its hardware alongside the TensorFlow framework~\cite{tensorflow2015-whitepaper}. 
Meanwhile, advancements in design-space exploration~\cite{dse} enabled architects and software engineers to evaluate thousands of hardware/software configurations to optimize performance, power, and cost. These successes underscored how close collaboration between computer architects and domain experts can yield dramatic gains. In fact, the end of general-purpose scaling forced teams to work hand-in-hand to create accelerators for AI, data analytics~\cite{inrdbms}, crypto~\cite{accel}, and more, effectively bringing co-design into a golden age of full-stack innovation~\cite{10876858}.
\subsection{Epoch 5 – Co-Design for Intelligence: Navigating the Redshift of Generative AI} 
We now stand at the cusp of a new epoch, driven by the rise of generative AI and extremely large-scale models. Transformer-based LLMs and other generative networks~\cite{team2023gemini} have exploded in size and capability, placing unprecedented demands on computing infrastructure that must be brutally efficient at deep scale to keep pace—a scenario reminiscent of Greg Papadopoulos’s description of "Redshift" applications, where growth outpaces traditional infrastructure development. In this ecosystem, the surging demand for AI not only revalidates but also expands the Redshift theory, representing one of the most extreme examples of a market in a state of redshift—a rapid, transformative acceleration that defies conventional limits.

In this epoch, AI is not merely an application; it has become the core software layer that redefines system design. Just as traditional software once redefined hardware capabilities, generative AI is now driving a profound shift where intelligence becomes an integral part of the computing stack. In this ecosystem, even the industry mindset for hardware design has shifted: hardware roadmaps now prioritize AI acceleration, even at the cost of traditional HPC needs. This evolution is reminiscent of how vertical integration in the 60s set the stage for today’s co-design approaches, yet the pace and scale of AI innovation demand a different design philosophy. 
Models have grown to hundreds of billions of parameters in just a few years and are projected to grow even further, quickly outpacing conventional hardware design principles. Unlike earlier embedded workloads, these models evolve rapidly every few months---or even weeks. Therefore, co-design approaches must be far more agile to keep up with the pace of innovation and have a fresh look at how we plan for a future where intelligence (models and data) is the key computing workload. We see hardware architects increasingly collaborating with ML researchers at every stage, from algorithm design to model training, to co-create solutions. In industry, this is evident through cross-disciplinary teams (architects, ML scientists, systems engineers) co-designing everything from new chips to datacenter-scale systems for generative AI.
It is important to note that this transition isn't merely technological; it also represents a market evolution.
As described in the “Crossing the Chasm” framework, generative AI now faces the critical challenge of moving from visionary (early adopters) to mainstream pragmatists (early majority). While a bandwagon effect might tempt companies and researchers to jump on the latest trend without thorough evaluations, it’s vital that we resist blindly following momentum. Instead, sustainable progress in this space demands that we carefully bridge the chasm with innovations rooted in effective hardware-software co-design, ensuring that breakthrough performance isn’t sacrificed for short-term hype.
\section{Co-Design Challenges in the Generative AI Era: Efficiency, Adaptability, and Complexity}
The generative AI boom thrusts hardware-software co-design into uncharted territory, bringing a distinct set of challenges:
\begin{itemize}[leftmargin=10pt]
    \item \textbf{Unprecedented Efficiency Demands:} Generative AI models push computing resources to their limits. For example, training Meta’s LLaMA-2 70B parameter model required roughly 1.72 million GPU hours on A100 GPUs~\cite{touvron2023llama}, with electricity cost soaring into the hundreds of thousands of dollars. Inference at scale faces similar constraints. Here, every percent of efficiency translates into massive savings, lower latency, reduced energy consumption, and the capacity to train larger models on practical budgets. This demands co-design that optimizes the entire stack—from alternative numerical precision and dynamic sparsity to enhanced data reuse and minimized memory transfers. Techniques like FlashAttention~\cite{dao2022flashattention}/FLAT~\cite{flat}, driven by GPU/TPU memory access patterns, and TPU’s bfloat16 support, co-evolved with ML software, underscore this point.
    \item \textbf{Need for Adaptability:} Traditional co-design delivered relatively fixed hardware tailored for near-stable workloads over multi-year cycles. In contrast, the generative AI landscape is a moving target. Hardware built for 2020-era transformers may soon be outdated if 2025-era models introduce new attention mechanisms or larger token contexts. Thus, adaptability must become a first-class design goal. We need software-defined hardware that combines the efficiency of custom silicon with the flexibility for post-silicon programmability. For example, designs featuring flexible dataflow architectures or programmable on-chip networks may provide the adaptability required to keep pace with rapid AI innovation.
    \item \textbf{Complexity, Memory, and Bandwidth Constraints:} Indeed, the challenge of co-design in this era extends beyond compute. LLMs demand vast amounts of memory (e.g. parameters, activations, and KV-caches), making memory capacity and bandwidth critical bottlenecks~\cite{gholami2024aimemorywall}. Effective co-design requires that memory systems be tightly integrated with model execution plans to ensure efficient proximity between data and compute units. Additionally, in distributed AI systems, orchestrating data-parallel and model-parallel strategies across GPUs or TPUs is essential for minimizing network overhead and achieving near-linear scaling. Reliability and correctness~\cite{sdc} become critical as these systems push hardware and software to their uncharted limits. The complexity of co-designing an entire AI pipeline, from model architecture down to transistors, also poses significant organizational and tooling challenges, necessitating robust cross-field collaboration.
\end{itemize}
In short, co-design for generative AI is a full-stack challenge that spans technical, organizational, and cross-domain fronts. The entire computing industry must reexamine its approach to hardware-software co-design and embrace continuous adaptation to keep pace with the relentless evolution of modern AI, recognizing the limitation of our current methods and remaining receptive to alternative strategies.

\section{The Hardware Lottery: How Hardware Shapes Winners and Losers in Research}
Compounding these issues is the hardware lottery, a phenomenon where a research idea succeeds not solely due to its inherent merits but because it aligns well with existing hardware and system software.
In other words, current hardware can unknowingly pick winners: algorithms that fit the hardware’s strengths flourish. The hardware lottery’s impact on research progress is double-edged. On one hand, it can accelerate progress by concentrating effort on approaches that are immediately tractable with existing hardware (as observed with deep learning riding the GPU wave at the AlexNet tipping point~\cite{krizhevsky2012imagenet}). 
On the other hand, it can hold back or delay potentially transformative ideas that don’t fit the mold of current machines. With the trend toward specialized AI accelerators, there’s a concern that we might be narrowing the field of viable algorithms too prematurely.
If chip design overwhelmingly favors Transformers~\cite{vaswani2017attention}, alternative AI paradigms might inadvertently be marginalized.
The hardware lottery paper argues that as hardware becomes more specialized, it “\emph{make[s] it increasingly costly to stray off of the beaten path of research ideas}”. In the context of generative AI, this could mean we double-down on one class of models (Transformers) because the hardware is optimized for it, while neglecting others (Text Diffusion) that might be better in the long run, but lack efficient hardware. The hardware lottery thus serves as a caution: \emph{co-design efforts must be careful not to inadvertently lock-in what is merely expedient, at the expense of what is optimal.}

\section{Toward a New Co-design Paradigm: Mitigating the Hardware Lottery’s Influence}
The evident path to mitigate the influence of hardware lottery is to broaden the exploration of both algorithms and hardware architectures.  The following principles, while seemingly apparent, are often overlooked and merit deeper investigation:
\begin{itemize}[leftmargin=10pt]
    \item \textbf{Reconfigurability and General-Purpose Flexibility:} To avoid prematurely committing to a “winner,” we must design hardware that accommodates a variety of algorithmic approaches (even if it means a modest sacrifice in efficiency!). 
    This requires a mindset that supports the integration of emerging, not-yet-fully-proven hardware features, developed in close collaboration with ML researchers, with more general-purpose, extensible architectures. For example, a chip could combine fixed accelerators for critical operations with a flexible engine that adapts to new computing idioms, thereby leveling the playing field for innovative algorithms.
    \item \textbf{Diversity in Hardware Architecture Research:} The current AI boom has already spurred a wave of startup activity in AI chips, generating a healthy variety of design approaches. Yet, we must continue to encourage and build on top of this diversity. Supporting a broad range of co-design projects, even those targeting non-mainstream workloads, reduces our reliance on a single outcome. History reminds us that pluralism in design, from the mainframe era to the personal computer revolution, has driven remarkable innovation. Additionally, academia should be encouraged to publish negative results and lessons learned, fostering a healthy environment where unconventional ideas can flourish without succumbing to the stifling grip of traditional publication standards.
    \item \textbf{Shorten the Iteration Cycle between Hardware and Algorithm:} One major challenge is that hardware development can take 2-3 years (not to mention maturing the supporting software stack), while ML algorithms often evolve in mere weeks or months. 
    By investigating rapid prototyping methodologies that shrink hardware iteration cycles to a matter of months (or possibly even weeks?) we can enable near-real-time co-design. This accelerated pace ensures that hardware, software, and algorithm development move in lockstep, ensuring none lag behind.
    \item \textbf{Build Co-design into the Culture of AI Research:} Historically, hardware considerations were often an afterthought for many ML researchers. To truly mitigate the hardware lottery, co-design must become an integral part of the research culture. Every proposal for a new model should include a consideration of its hardware requirements, and every hardware innovation should clearly articulate the new algorithms (or computing idioms) it supports. Interdisciplinary collaboration—integrating architects within ML teams and vice versa—can bridge the gap between the two fields and ensure that insights flow freely without any socio-emotional judgments from either side, and that they can live together happily ever after.
\end{itemize}
Together, these principles, when combined with a renewed focus on education pedagogy in computer architecture, can drive a shift in design philosophy. Once, we enjoyed the comfort of Moore’s Law and an abundance of transistors, which allowed us to take a more relaxed approach to hardware design. Today, if I may say, the discomfort of unprecedented AI demands challenges us to rethink our approach, making it essential to instill a cross-stack thinking mindset in both students and researchers. By embracing flexibility, diversity, rapid iteration, and a culture of tight collaboration, we can diminish the constraints of the hardware lottery and shape the future of computing. I speculate that as the impact of hardware-software co-design becomes increasingly pronounced, it may well be doubling generative AI performance roughly every 12 months, outstripping the gains once driven by Moore’s Law.

\section*{Acknowledgements}
I want to thank my colleagues and collaborators for their valuable feedback and insights while developing the ideas for this paper.
Special thanks to Cliff Young (Google DeepMind), Suvinay Subramanian (Google), and Michael D. Smith (Harvard) for sharing their anecdotes about co-design with me, which greatly helped to revise the initial version of this draft. 
In addition, I appreciate the valuable feedback from Herman Schmit (Google) and Derek Lockhard (Google DeepMind). 
This post draws significant inspiration from the ideas presented in~\cite{6172642,1193227,10876858,hennessy2017computer,sigarch_history_part1,sigarch_history_part2}.

\bibliographystyle{plain}
\bibliography{paper}

\end{document}